\documentclass[numreferences]{kluwer}    
\usepackage{klucite}
\usepackage{epsfig}
\newdisplay{guess}{Conjecture}

\begin{document}
                     
\begin{article}
\begin{opening}         
\title{SZ scaling relations in Galaxy Clusters: results from
            hydrodynamical N--body simulations} 
\author{Ant\'onio J. C. \surname{da Silva}}  
\runningauthor{Ant\'onio da Silva}
\runningtitle{SZ scaling relations in galaxy clusters from
            hydrodynamical simulations}

\institute{LAOMP, 14, Av. Edouard Belin, 31400 Toulouse, France}
\begin{abstract}
Observations with the SZ effect constitute a powerful new tool for
investigating clusters and constraining cosmological
parameters. Of particular interest is to investigate how the SZ signal
correlates with other cluster properties, such as the mass,
temperature and X-ray luminosities. In this presentation we quantify these
relations for clusters found in hydrodynamical simulations of large scale
structure and investigate their dependence on the effects of radiative
cooling and pre-heating. 
\end{abstract}

\end{opening}           

\section{Introduction \label{intro}}  

Galaxy clusters are the largest gravitationally bound objects in the
Universe. They have typical masses of $10^{14}-10^{15}h^{-1}$~M$_\odot$
and contain hundreds of galaxies within radius of a few Mpc. 
The intra-cluster medium (ICM) is filled with hot ionized gas,
typically at temperatures 1--15 keV, which produces strong X-ray emission and
causes spectral distortions in the CMB spectrum via the
Sunyaev--Zel'dovich (SZ) effect (see Ref.~\cite{sunyaev:1972}). 

Numerical simulations indicate that the non-baryonic dark matter component
in clusters, which is the dominant fraction of their mass, is remarkably
self-similar for systems in  approximate state of equilibrium, see e.g.
Ref.~\cite{navarro:1997b}.
However, the baryonic gas component does not share the same degree of
self-similarity. This is more evident from observations of the
$L_X$--$T$ relation in clusters, which is much steeper than is
predicted by simple self-similar scaling laws, specially
for low-mass systems 
(see e.g. Refs.~\cite{edge:1991,xue:2000}). 
This deviation from self-similarity has been interpreted as due to 
non-gravitational processes, such as radiative cooling and heating,
that raise the entropy of the gas, see e.g. Ref.~\cite{ponman:1999}.

The purpose of this study is to investigate the correlation between
intrinsic properties of clusters found in simulations, and to 
compare them with analytical scaling laws. 
We use high-resolution hydrodynamical simulated clusters to assess the impact
of radiative cooling and pre-heating on these scaling relations. 
Here we focus only on scalings involving the SZ integrated 
signal, $Y$, at redshift zero. The effects  
on the mass--temperature and X-ray scaling relations have already been 
analyzed in detail in Refs.~\cite{muanwong:2001,muanwong:2002}
for this same set of simulations. 
A more detailed analysis of the SZ scaling relations can be found in 
Ref.~\cite{dasilva:2003}, whose results supersede those presented 
at this conference.

\section{Scaling relations in Clusters \label{szscal}}
The quantity we want to study is the total thermal SZ flux
density received from a cluster. This is defined as the integral of the 
thermal SZ specific intensity (described by the $y$ Comptonization 
parameter) over the solid angle occupied by the source 
in the sky. More exactly, we want to investigate 
how the frequency-independent SZ flux,   
%
\begin{equation}
Y\equiv \int {y\, d\Omega}=d_A^{-2}\int {y\,dA}=
{ {k_B\sigma _{\rm T}} \over {mc^2}}\,d_A^{-2}\,
\int _{V}\, T_{{\rm e}} n_{{\rm e}} \, dV\,,
\label{cp6_4}
\end{equation}
correlates with other cluster properties. This is known as the SZ
integrated $Y$-flux or $Y$-luminosity. 
Note that $d_A$ is the angular diameter distance, $dA=d\Omega \,d_A^{-2}$
is the sky projected area of the source and   
the last integral is performed over the volume of the
cluster.
%
Since $y$ is dimensionless, $Y$ has dimensions of a solid
angle. 
Equation~(\ref{cp6_4}) indicates that $Y$ is proportional to the mean electron
temperature, $\langle T_{\rm e} \rangle$, and the total cluster mass, $M$, 
%
\begin{equation}
Y\propto f_{\rm gas}\, \langle T_{\rm e} \rangle \, M\,d_A^{-2},
\label{cp6_5}
\end{equation}
where $f_{\rm gas}$ is the gas mass fraction of the cluster. 
Note that $\langle T_{\rm e} \rangle$ is the mass (or particle) weighted
temperature, 
which may differ from the {\it emission-weighted} temperature 
measured in X-ray observations.

The dependence of the X-ray bolometric luminosity on mass and
temperature follows from its definition,
%
\begin{equation}
L_X =\int_{V}\, \frac{\rho_{\rm gas}^2}{(\mu m_{\rm p})^2} \Lambda (T)\, dV\,.
\label{cp6_8}
\end{equation}
The integral is over the volume of the cluster, 
$\mu m_{\rm p}\simeq 1\times 10^{-24}$ is the mean mass per particle, 
$\Lambda (T)$ is the gas cooling function and 
$\rho_{\rm gas}$ is the gas density. Assuming the gas is well described by an 
isothermal temperature profile one obtains, 
%
\begin{equation}
L_X \propto f_{\rm gas}^2\, M\,\Delta _{\rm
c}\,\rho _{\rm crit}\,\Lambda (T)\,
\propto f_{\rm gas}^2\, M\,\Delta _{\rm
c}\,\rho _{\rm crit}\,T^{1/2}\,,
\label{cp6_9}
\end{equation}
where the last step results from assuming that the
bolometric luminosity in clusters is dominated by bremsstrahlung
emission, $\Lambda \propto T^{1/2}$.

As suggested by the virial theorem the mass and the temperature of a
cluster are tightly correlated quantities. At the virial radius these 
are related by the virial relation, $k_{\rm B}T\propto GM/r$.
The mass enclosed within this radius can be written as 
$M=4\pi \,r^3\,\Delta _{\rm c}\,\rho _{\rm crit}/3$, 
where $\Delta _{\rm c}$ is the overdensity contrast between the mean cluster
density within $r$ and the critical density, $\rho _{\rm crit}$. 
Using this expression one obtains 
%
\begin{equation}
T\propto M^{2/3}\,\left( {\Delta _{\rm
c}\,\rho _{\rm crit} }\right) ^{1/3}\,
\propto \left( {\frac{\Delta _{\rm c}}{\Omega(z)} } \right)^{1/3}
M^{2/3}\, (1+z) \,.
\label{cp6_7}
\end{equation}

Using Eq.~(\ref{cp6_7}) into Eqs.~(\ref{cp6_5}) and (\ref{cp6_9}),
with $\langle T_{\rm e} \rangle \sim T$, one can
derive general scaling relations for the SZ and X-ray luminosities 
as a function of cosmology, redshift, and the cluster mass (or temperature). 
According to these, clusters are expected to present the following
interdependencies of mass, temperature, X-ray luminosities and
integrated SZ fluxes at redshift zero:
%
\begin{equation}
T\propto M^{2/3}\,,\quad \,\,
L_X\propto  M^{4/3}\propto T^{2}\,,\quad \,\,
Y\propto  M^{5/3}\propto T^{5/2}\propto L_X^{5/4}\,.
\label{cp6_12}
\end{equation}
These expressions coincide with the scaling relations
predicted by the self-similar model \cite{kaiser:1986}, 
which is expected to describe well the correlation between the above 
properties if shock heating from gravitational collapse is the dominant 
mechanism driving the gas evolution. 

\section{Simulations, Cluster identification and catalogues}
We present results from three simulations of a single 
$\Lambda$CDM cosmology with density parameter, $\Omega_{{\rm m}}=0.35$, 
cosmological constant, $\Omega_{\Lambda}=0.65$, Hubble parameter $h=0.71$,
baryon density, $\Omega_{{\rm B}} h^2=0.019$, shape parameter of the 
CDM power spectrum, $\Gamma=0.21$, and normalization $\sigma_8=0.9$.
We use $160^3$ particles of each gas and dark matter in a 
box of side $100 \, h^{-1} {\rm Mpc}$, which gives particle masses of
$m_{\rm gas}=2.6\times 10^{9} \, h^{-1} {\rm M_{\odot}}$ and
$m_{\rm dark}=2.1 \times 10^{10} \, h^{-1} {\rm M_{\odot}}$ respectively. The 
gravitational softening at redshift zero was $25\,h^{-1} {\rm kpc}$.
The simulations were carried out as part of the Virgo Consortium 
program, using a parallel version of the {\tt hydra} N-body/hydrodynamics 
code (see Refs.~\cite{couchman:1995,pearce:1997}).

The first of the three simulations was performed without any additional
heating or cooling mechanisms; we refer to this as the `non-radiative'
simulation. The second simulation included a model for radiative cooling
using the method described in Ref.~\cite{thomas:1992}, except that we adopted
the cooling tables of Ref.~\cite{sutherland:1993}
and a global gas metallicity evolving as $Z=0.3(t/t_0) Z_{\odot}$, where 
$Z_{\odot}$ is the solar metallicity and $t/t_0$ is the cosmic time in units 
of its present value. Our third simulation, which also includes cooling, was 
performed in order to study the effects of pre-heating the gas at high 
redshift. We considered an energy injection of 0.1 keV per gas particle at
redshift four for this run. Note that this is less energy injected per 
particle than that used in the pre-heating run of 
Ref.~\cite{dasilva:2003} (1.5 keV).
We used identical initial conditions to enable 
direct comparisons between clusters forming in the three runs. For more 
details see Ref.~\cite{dasilva:2001b}.

In simulations clusters are identified with a spherical overdensity
group finder algorithm, as described in 
Ref.~\cite{muanwong:2001}.
The identification process consists of
several steps. The first is to create a minimal-spanning tree of all
dark-matter particles whose density exceeds $\delta=178\times
\Omega^{-0.55}(z)$ times the mean dark-matter density in the box. 
The next step is to split the minimal-spanning tree into clumps of 
particles using a maximum linking length equal to $0.5\delta^{-1/3}$
times the mean inter-particle separation.
Then we grow a sphere around the centre of the clump until
the enclosed mean density is $\Delta $ times larger then the
comoving critical density. 
Because the density parameter of the real universe is not known, we
have chosen to average the properties within an isodensity
contour of $\Delta=200$, corresponding to a radius $r_{200}$.  
A final cluster catalogue is 
produced by selecting objects whose total mass is equivalent to at
least 500 particles of gas and dark matter within this radius, and
whose centre is not located within a more massive cluster. In this
way our catalogues (available online from {\tt virgo.sussex.ac.uk}) 
are complete in mass down to 
$M_{\rm lim}\approx1.18\times10^{13}h^{-1}M_\odot$. 

For each cluster the catalogues contain several entries representing
different cluster properties. Relevant for this study are the bolometric 
luminosity, 
$L_{\rm bol} = \sum \, m_i \, \rho_i \,\Lambda(T_i,Z)/(\mu m_{\rm p})^2$,
the emission-weighted temperature, 
$T_{\rm ew} = \sum \, m_i \, \rho_i \Lambda(T_i,Z) \, T_i/ \sum \, m_i \, \rho_i \Lambda(T_i,Z)$, 
the mass-weighted temperature
$T_{\rm mw} = {\sum \, m_i \, T_i/ \sum \, m_i }$,
and the hot gas mass fraction,
$f_{\rm gas} = \sum \, m_i /M$.
In these expressions, summations run over gas particles with temperatures 
above $10^5\,(1+z)^2$ K (i.e. the hot intra-cluster gas that gives rise to 
the SZ and X-ray emission) within a radius $r_{200}$. M is 
the total mass, i.e., the sum of the mass of all particles, including both 
baryons and dark matter, within the same radius. The quantities, 
$m_i$, $T_{i}$, $\rho _{i}$ and $\Lambda(T_i,Z)$ are respectively the mass,
temperature, density and emissivity (cooling function) of the
particles. 

The intrinsic SZ luminosity, defined as $Y^{\rm int}=Y\,d_A(z)^2$, can
then be computed from the cluster catalogues as, 
\begin{equation}
Y^{\rm int} =
\frac{\sigma_{\rm T}\,k_{\rm B}}{m_{\rm e}c^2}\,\frac{(1+X)}{2m_{\rm p}}\,
\sum\limits_{i \in {\rm hgas}} \, m_i \, T_i
= \frac{\sigma_{\rm T}\,k_{\rm B}}{m_{\rm e}c^2}\,\frac{0.88}{m_{\rm p}}\, 
f_{\rm gas}\, T_{\rm mw}\, M\,, 
\label{cp6_18}
\end{equation}
where we assume a hydrogen mass fraction of $X=0.76$. Since the $y$-parameter 
is dimensionless we will quote $Y^{\rm int}$ in units of $(h^{-1} {\rm
Mpc})^2$. 

\section{Results: scaling relations at redshift zero}
\begin{figure}
\centering \leavevmode\epsfysize=6.2cm \epsfbox{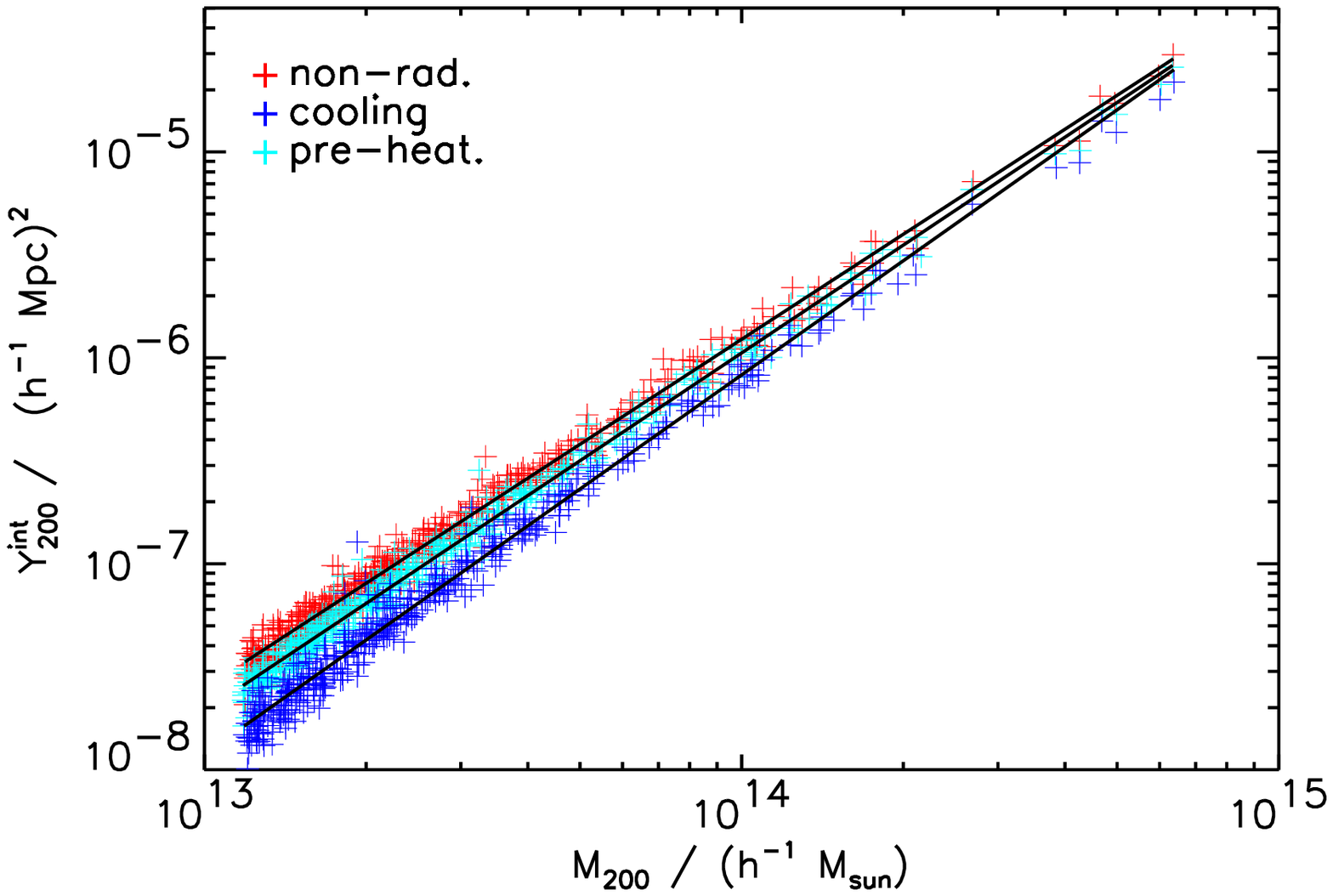}\\ 
\vspace{-0.5cm}
\centering \leavevmode\epsfysize=6.2cm \epsfbox{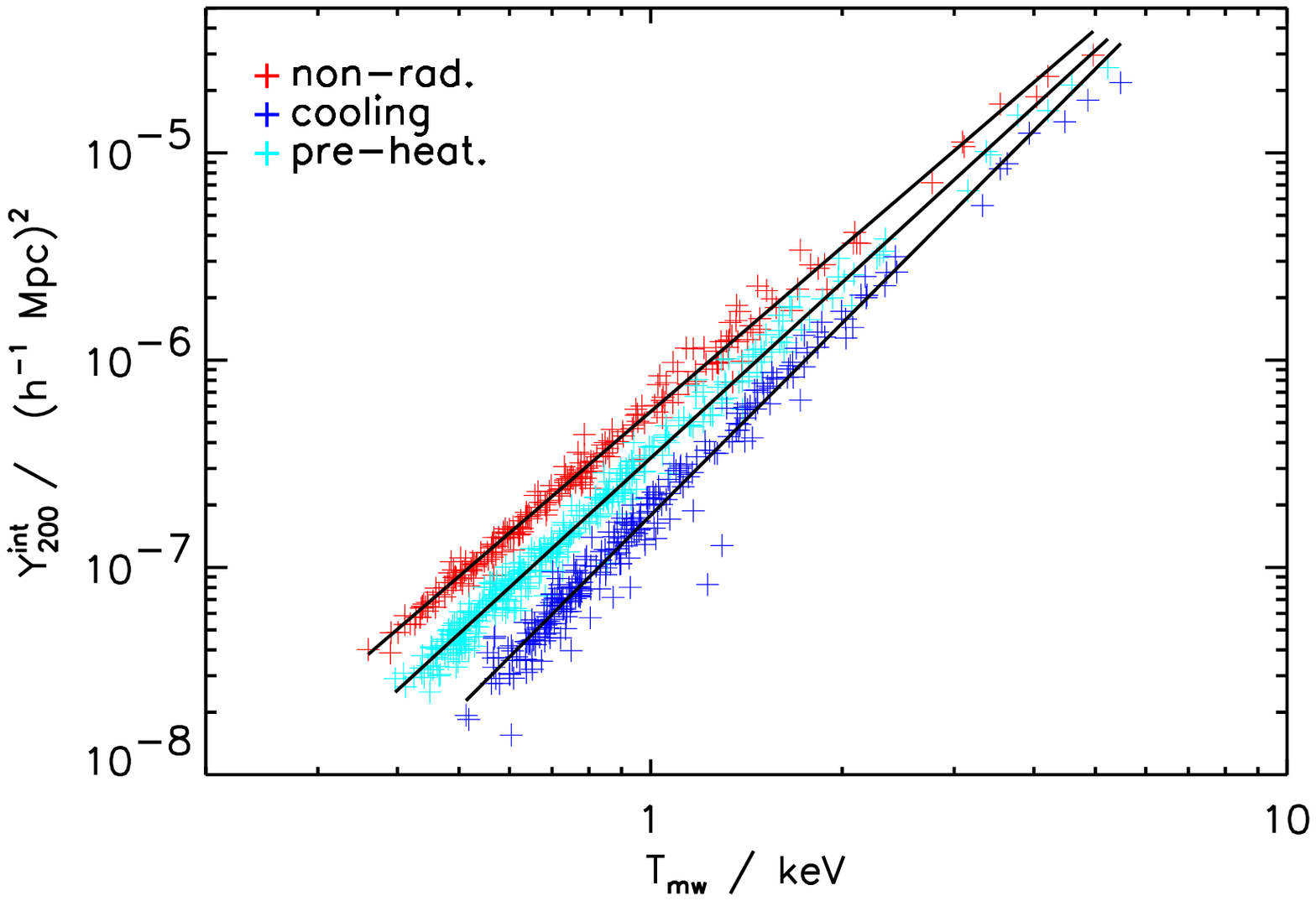}\\
\vspace{-0.5cm}
\centering \leavevmode\epsfysize=6.2cm \epsfbox{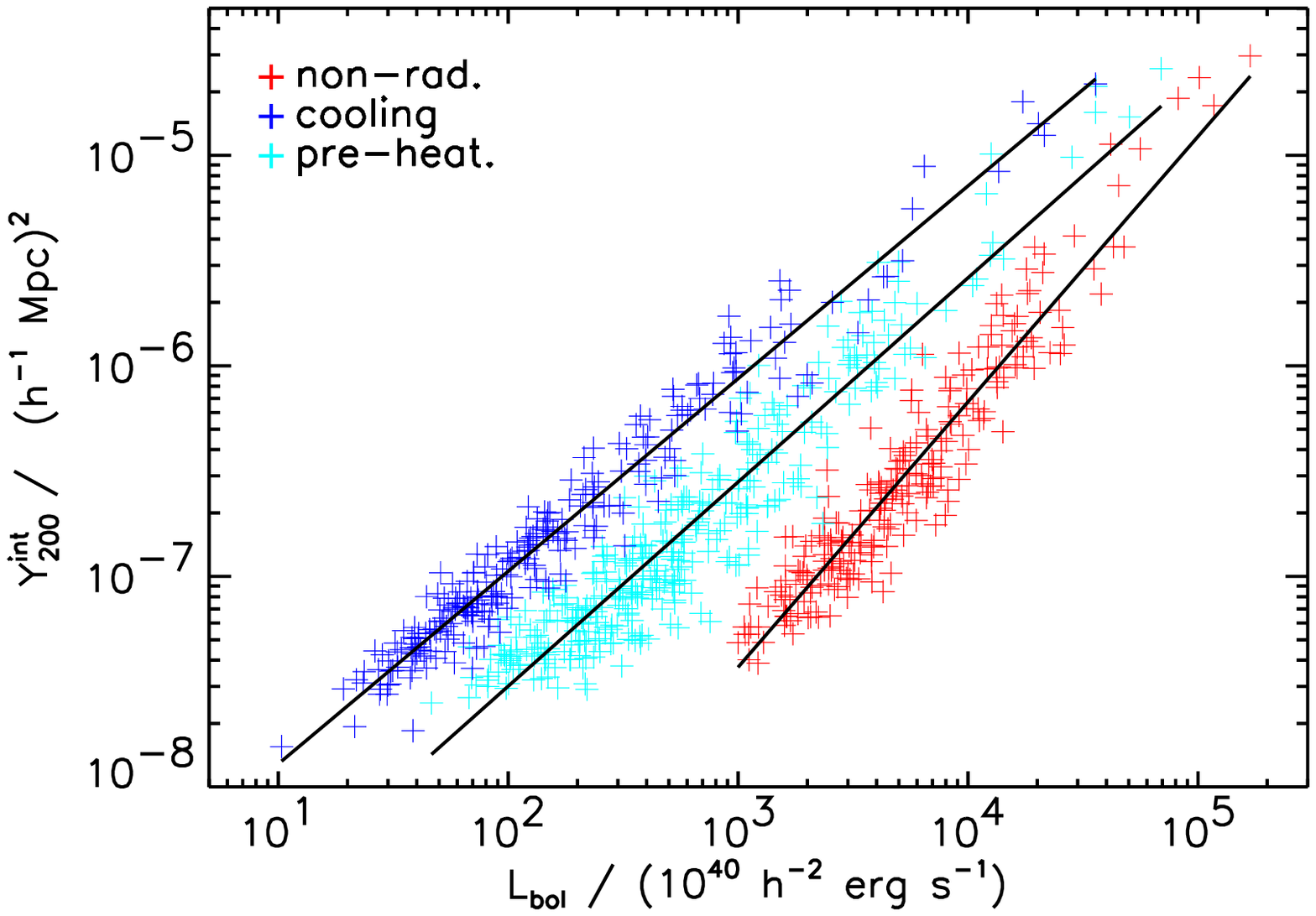}\\
\caption[SZ scaling relations for clusters at redshift zero from
non-radiative, cooling and pre-heating simulations of the $\Lambda
$CDM cosmology.  ]{SZ scaling relations for clusters at redshift zero
from the non-radiative (gray crosses), cooling (dark gray crosses) and
pre-heating (light gray crosses) simulations of the $\Lambda $CDM cosmology.
From top to bottom we have the scalings of the intrinsic $Y$-flux with
mass, mass-weighted temperature and X-ray bolometric luminosity,
respectively. The solid lines are power-law fits to the
distributions.}
\label{fig:scale}
\end{figure}

In Figure~\ref{fig:scale} we plot the intrinsic SZ-flux against mass 
(top panel), mass-weighted temperature (centre panel) and X-ray luminosity 
(bottom panel), for clusters found in simulation boxes at redshift 
zero. All properties are calculated within $r_{200}$. 
For the plots in this paper we adopt the following convention: gray
crosses represent clusters from the non-radiative simulation, whereas
dark gray and light gray crosses illustrate the properties of clusters
in the cooling and pre-heating runs.  To ensure completeness in
temperature and luminosity we have trimmed the original cluster
catalogues in the two bottom plots by selecting only clusters with
emission-weighted temperatures above 0.35, 0.75 and 0.55 keV for the
non-radiative, cooling and pre-heating simulations, respectively.
This permits to avoid the introduction of artificial trends on the 
derived fittings for the $Y^{\rm int}_{200}-T_{\rm mw}$ and 
$Y^{\rm int}_{200}-L_{\rm bol}$ relations.
%

The plots show that $Y^{\rm int}_{200}$ is tightly correlated with 
mass and the mass-weighted temperature in all runs. 
The correlation between the SZ integrated flux and the X-ray
luminosity is also strong, but shows 
significantly more scatter. This is because the X-ray emission is 
sensitive to the gas distribution and temperature, whereas 
$Y^{\rm int}_{200}$ is essentially a function of the total thermal 
energy of the hot gas mass within $r_{200}$. In each plot, the solid 
lines are power-law best fits to distribution of clusters in 
each simulation. 
For the $Y^{\rm int}_{200}-M_{200}$ correlation we find the following 
best-fit parameters for each simulation run:

Non-radiative:~~~$Y^{\rm int}_{200}=1.2\times 10^{-6}\,\left( {
M_{200}/M_{14}} \right)^{1.7}\,\left( { h^{-1}\,{\rm Mpc} } \right)^2$

Cooling:~~~~~~~~~~~$Y^{\rm int}_{200}=8.3\times 10^{-7}\,\left( { 
M_{200}/M_{14}} \right)^{1.8}\,\left( { h^{-1}\,{\rm Mpc} } \right)^2 $

Pre-heating:~~~~~~$Y^{\rm int}_{200}=1.0\times 10^{-6}\,\left( {
M_{200}/M_{14}} \right)^{1.7}\, \left( { h^{-1}\,{\rm Mpc} } \right)^2 $,

\noindent{where $M_{14}=10^{14}\,h^{-1}\,{\rm M}_\odot$.
The slope from the non-radiative run reproduces well
the value 5/3 predicted by the self-similar scaling Eq.~(\ref{cp6_12}). 
The cooling simulation presents a steeper
slope and lower normalization then the other two runs.
This implies significant differences in  
$Y^{\rm int}_{200}$, particularly for low-mass systems. For example, at
$M_{200}\sim 10^{13}\,h^{-1}\,{\rm M}_\odot$, sources in the cooling
simulation are about two times less strong then their non-radiative
counterparts. 
At $10^{14}\,h^{-1}\,{\rm M}_\odot$ the relative 
difference is still of about 50 percent, but gets much reduced at higher
masses.}

The differences between runs can be explained in terms of the changes
caused by our models of cooling and pre-heating in the hot gas
fraction and mass-weighted temperature of clusters. The inclusion of
cooling in both radiative simulations lowers the hot gas mass
fraction and tends to increases the emission-weighted
temperature due to the removal of cooled material 
from the hot gas phase. The increase in temperature results from the infall 
of high-entropy gas which replaces the cooled material forming at the
central regions of clusters. 
As can be inferred from Eq.~(\ref{cp6_18}), these are
competing effects for the same cluster mass. The overall effect on $Y^{\rm
int}_{200}$ is dominated by the decrement of the hot gas mass
fraction, which shifts clusters downward in the $Y^{\rm
int}-M$ plane. The effect is less strong in the pre-heating run,
because particles were heated up by the energy injection mechanism and
less amount of cooled material is able to form. The decrement of
$Y^{\rm int}_{200}$ is also stronger in low-mass systems, which explains
the steepening of the relations.

\begin{figure}[t]
\begin{center}
\epsfysize=6.2cm \epsfbox{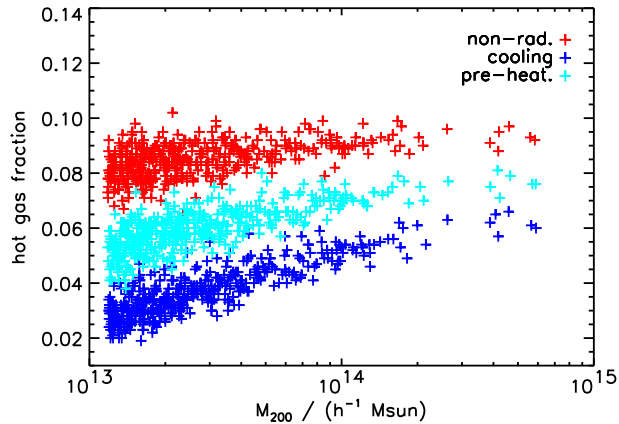}
\caption[Fraction of hot gas within $r_{200}$ for clusters in the
non-radiative cooling and pre-heating simulations.]
{
Fraction of hot gas within $r_{200}$ for clusters in the non-radiative
(gray crosses) cooling (dark gray crosses) and pre-heating (light gray crosses)
simulations. 
}
\label{fig:fhgas_b} 
\end{center}
\end{figure}

In Figure~\ref{fig:fhgas_b} we plot $f_{\rm gas}$ for all clusters in
non-radiative, cooling and pre-heating catalogues at redshift zero. 
We see that clusters in the cooling simulation have the lowest hot gas
fractions. For low-mass systems these are reduced by about 2.5 times
when compared with clusters in the non-radiative simulation. The
fraction of hot gas in clusters of the pre-heating run is also
substantially reduced, but less than in the cooling case. In both
radiative simulations, lower mass systems show lower amounts of hot
gas, which implies a lower $Y^{\rm int}_{200}$. 
In the cooling simulation, this is because cooling becomes more
efficient in low-mass systems, where particles have lower
temperatures and higher cooling rates due to collisional excitation
cooling from neutral elements. 
In the pre-heating simulation, low-mass systems may also reduce their
hot gas fraction in this way. This process is less efficient
then in the cooling case because the energy injection rises the
temperature of the particles and reduces the contribution from
collisional excitation cooling.  
Pre-heating can also reduce the amount of hot gas in clusters in a
different way. Instead of converting hot gas into cooled material,
pre-heating has the effect of heating up and expelling gas from the
central regions of clusters. In low-mass systems this may cause some
of the hot gas to be expelled from the haloes, which in turn implies
lower $f_{\rm gas}$ at lower masses. We find that the amount of
baryonic gas which is converted into cooled material in clusters of
the cooling and pre-heating simulation can be as high as 50 and 30
per cent, respectively.

The correlation between $Y^{\rm int}_{200}$ and the mass-weighted
temperature is also easy to understand. For the same $T_{\rm mw}$,
Eq.~(\ref{cp6_18}) tells us that the $Y$-luminosity is simply
proportional to the hot gas mass of the cluster. 
As a consequence, for
a given temperature, clusters of the cooling and pre-heating
simulations present lower $Y^{\rm int}_{200}$. 
The effect is again stronger for low-mass systems and for clusters in
the cooling simulation. Clusters 
in the radiative runs also present higher temperatures for the same
mass. The combination of these two effects causes clusters to shift in
the direction of high temperatures and lower $Y$-fluxes in the
$Y^{\rm int}_{200}-T_{\rm mw}$ plane. The best-fit parameters found
for the $Y^{\rm int}_{200}-T_{\rm mw}$ relation are:

Non-radiative:~~~$Y^{\rm int}_{200}=5.8\times 10^{-7}\,\left( {T_{\rm mw}
/{\rm keV} } \right)^{2.6}\,\left( { h^{-1}\,{\rm Mpc} } \right)^2 $

Cooling:~~~~~~~~~~~$Y^{\rm int}_{200}=1.8\times 10^{-7}\,\left( {T_{\rm mw}
/{\rm keV} } \right)^{3.0}\,\left( { h^{-1}\,{\rm Mpc} } \right)^2 $

Pre-heating:~~~~~~$Y^{\rm int}_{200}=3.4\times 10^{-7}\,\left( {T_{\rm mw}
/{\rm keV} } \right)^{2.8}\,\left( { h^{-1}\,{\rm Mpc} } \right)^2 $

\noindent{The slopes of the 
relations derived from the radiative simulations are steeper than those 
found in the non-radiative run. This is due to the higher temperatures and
lower gas masses of clusters in these simulations. Again the
non-radiative simulation is in good agreement with the slope 5/2
predicted by the self-similar scaling.}

\begin{figure}[t]
\begin{center}
\epsfysize=7.75cm \epsfbox{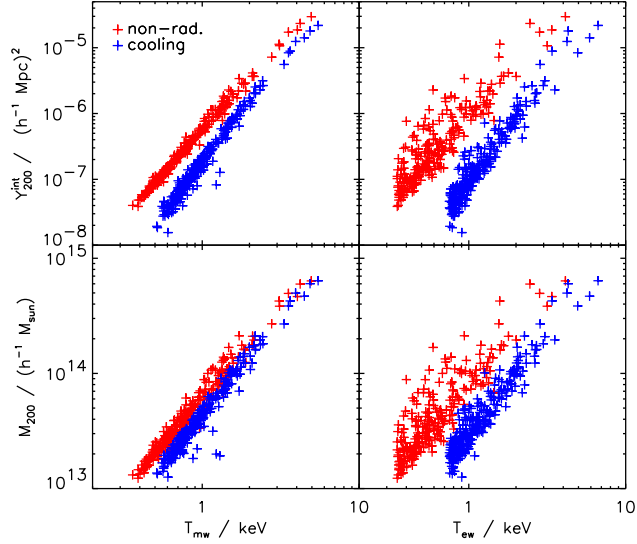}
\caption[The integrated SZ flux-temperature and mass--temperature relations in
clusters from the non-radiative and cooling simulations.]  
{The top panels show the $Y^{\rm int}_{200}$ against the mass-weighted
(left panel) and X-ray emission-weighted (right panel) temperatures
(respectively $T_{\rm mw}$ and $T_{\rm ew}$) for clusters in the
non-radiative (gray crosses) and cooling (dark gray crosses)
runs. The bottom panels show mass--temperature correlations
for the same set of clusters.
}
\label{f:ym-t4}
\end{center}
\end{figure}

In the top panels of Figure~\ref{f:ym-t4} we plot $Y^{\rm int}_{200}$
against the mass-weighted (left panel) and X-ray emission-weighted
(right panel) temperatures for clusters in the non-radiative and
cooling simulations. These show that the integrated SZ signal presents
a much tighter correlation with the mass-weighted temperature than with the
emission-weighted temperature. This result confirms expectations,
since $T_{\rm ew}$ is weighted by density and temperature and is
therefore sensitive to substructures and clumping of the gas inside
the cluster. 
The bottom panels of Figure~\ref{f:ym-t4} show the correlation between
mass and temperature for the same set of clusters displayed in the top
panels. Again, the correlation between $M_{200}$ and $T_{\rm mw}$
presents significantly less scatter than the correlation of $M_{200}$
with $T_{\rm ew}$. This is evidence of the potential of the SZ effect
to study the properties of clusters. Future high-resolution SZ
observations may provide important constraints on cluster evolutionary
theories. Good calibrations of the observed $Y-T$ and $Y-M$
relations may also provide alternative robust estimates of temperature and
mass in clusters.  

We end the analysis by reporting on the $Y^{\rm int}_{200}-L_{\rm
bol}$ correlation. In the
radiative runs, the high entropy gas which replaces the cooled material
which forms in these simulations is less dense and
hotter then the gas in the non-radiative simulation. These variations
in temperature and density have opposite effects on the X-ray
luminosity, which is a slowly-varying function of temperature, but it is
proportional to the square of the gas density, see
Eq.~(\ref{cp6_8}). As a result, the change in density wins the
competition and the luminosity decreases with cooling. The inclusion
of pre-heating reduces the amount of cooled material and increases the gas
temperature. This implies higher luminosities in the pre-heating run
then in the cooling simulation, but lower luminosities than in the
non-radiative run. The combination of these effects with the decrease
of $Y^{\rm int}_{200}$ explains the substantial shifts observed in
the $Y^{\rm int}_{200}-L_{\rm bol}$ relations of
Figure~\ref{fig:scale} in the pre-heating and cooling simulations. The
best-fit parameters for the distribution of clusters 
in the $Y^{\rm int}_{200}-L_{\rm bol}$ plane are:

Non-radiative:~~~$Y^{\rm int}_{200}=6.3\times 10^{-12}\,\left( {
L_{\rm bol}/L_{40} } \right)^{1.3}\,\left( { h^{-1}\,{\rm Mpc} } \right)^2$  

Cooling:~~~~~~~~~~~$Y^{\rm int}_{200}=1.6\times 10^{-9}\,\left( {
L_{\rm bol}/L_{40} } \right)^{0.9}\,\left( { h^{-1}\,{\rm Mpc} } \right)^2$ 

Pre-heating:~~~~~~$Y^{\rm int}_{200}=3.2\times 10^{-10}\,\left( {
L_{\rm bol}/L_{40} } \right)^{1.0}\,\left( { h^{-1}\,{\rm Mpc} } \right)^2$, 

\noindent{where $L_{40}=10^{40} h^{-2} {\rm erg\, s}^{-1}$.
These show considerable differences of slope and
normalization. The slope in the non-radiative run is close to 1.25, 
which is the value
predicted by the self-similar scaling. The
radiative runs show similar slopes, which are less steep than in the 
non-radiative case. As in the case of the $Y^{\rm
int}_{200}-T_{\rm ew}$ relation, the $Y^{\rm int}_{200}-L_{\rm bol}$
relation shows significant scatter due to the sensitivity of the X-ray
emission to the density and temperature distribution of the hot gas.}

\section{Conclusions}

We have studied the effect of our models of cooling and
pre-heating on the SZ cluster population by identifying clusters in
the simulation boxes and computing their
characteristic properties. We correlated the SZ luminosity with other
cluster properties and derived scaling relations at redshift zero. The
non-radiative simulation reproduces well the scalings predicted 
by the self-similar model, whereas the inclusion of cooling and
pre-heating generally changes the slope and normalization of the
scaling relations for those runs. The integrated $Y$-signal is found
to be tightly correlated with the mass-weighted
temperature and total mass of the cluster.

The author wishes to thank A. Liddle, P. Thomas, S. Kay and O.
Muanwong for many fruitful discussions and acknowledges the use of the
computer facilities of the Astronomy Unit at Sussex, UK, and CALMIP at
Toulouse, France. The simulations used were carried out on a Cray-T3E
at EPCC as part of the VIRGO Consortium collaboration.

%

\bibliographystyle{klunum.bst}
\bibliography{phd}

\end{article}
\end{document}